\documentclass{KEPASSA2024}
\usepackage{graphicx} % si figures
\title{Gradient based reachability analysis for goal-oriented guidance. Application to Eros proximity operations}
\author[1]{Antonio Rizza\thanks{Email: antonio.rizza@polimi.it}}
\author[1]{Francesco Topputo\thanks{Email: francesco.topputo@polimi.it}}
\affil[1]{Polytechnical University of Milan, Department of Aerospace Science and Technology, Via La Masa 34, 20156, Milan, Italy}
\newcommand{\STM}[2]{ \vec{\Phi}(#1,#2) }

\let\vec\mathbf

\begin{document}
	\maketitle

	%\begin{abstract}
	%Goal-oriented guidance strategies for small bodies proximity exploration enable high levels of autonomy to achieve cost-effective safe and reliable operations. Abstract reachability analysis performed on the control domain is a key tool in this framework but its computational burden prohibits on-board implementation. A trade-off of different gradient based approaches to explore the sets is proposed in this paper comparing results in terms of accuracy and computational effort. Results indicate that a combination of sample based and gradient based techniques outperforms classical heuristic methods. Overall, this work presents a step towards on-board goal-oriented autonomous guidance capability for small bodies proximity operations.
	%\end{abstract}
	
	\section{Introduction}\label{sec: introduction}
	
	The recent growing interest in small solar system bodies such as asteroids and comets for scientific inspection, exploitation of resources, and planetary defense reasons is pushing the development of innovative engineering solutions to better investigate these celestial bodies. Many robotic missions successfully performed scientific operations in proximity of asteroids, comets, and minor planets \cite{scheeres2016orbital}. Current approaches for autonomous proximity operations often relies on tracking a reference trajectory previously designed and optimized on-ground. This approach allows for no replanning capability, struggling in facing unforeseen events. Moreover, the satisfaction of specific observation and mapping requirements is often verified a-posteriori and not fully taken into account during the design phase. An innovative concept developed in recent years is proposing a paradigm shift towards autonomous goal-oriented approaches \cite{de2010taking}, providing the probe with high-level tasks and enabling replanning capability on-board. A sample based abstract reachability analysis performed in the control domain is proposed as a way of planning impulsive maneuvers within a receding horizon model predictive control framework in different works \cite{surovik2016autonomous,capolupo2017heuristic,rizza2023goal}. While being very flexible to different mission scenarios and observation requirements, this approach presents a few limitations mainly due to the computational cost of the reachable set. In fact, despite exploiting an heuristic refinement technique this approach still requires massive trajectories propagation and potentially specialized hardware to be performed on-board. A gradient-based variation of this approach is presented in this work facilitating the exploration of the reachable map converging to an optimal or sub-optimal solution with a lower computational effort. This abstract is divided as follows: Sec.\ref{sec:problem statement} presents the problem statement and methodology, Sec.\ref{sec: scenario definition} defines a test case in which the approach is validated, Sec.\ref{sec: results} shows a performance comparison among different numerical techniques to explore the set identifying the best approach.
	
	\section{Problem statement}\label{sec:problem statement}
	In the context of this work, the goal-oriented guidance problem consists in mapping the observation requirements into a series of discrete observation regions $\Omega_i$ defined in an abstract  space \cite{rizza2023goal}, and finding a series of impulsive control actions $\vec{u}_{0j}$ and the associated maneuvering epoch $t_{hj}$ over a given time horizon $T_{h,j}$ such that the spacecraft crosses the highest number of regions. This is achieved solving one trajectory arc at the time and imposing the observation requirements in a strong form including them in the objective function $J\left(\vec{u}_{0,j},t_{h,j}\right)$ to be maximized. The optimization problem is thus formulated as: 
	\begin{equation}
		\begin{aligned}
			&\max_{\vec{u}_{0,j},t_{h,j}} J\left(\vec{u}_{0,j},t_{h,j}\right) \text{ , such that:}\\
			&J = \int_{t_{0,j}}^{t_{h,j}}  \sum_{i = 1}^{n_{\Omega_i}}  \omega_i \left(\vec{y}_i \left(\tau\right)\right)  + V_C \left(\Vec{x} \left( \tau \right)\right)d\tau\\
			&\vec{y}_i\left(t\right) = \vec{M}_i \left( \Vec{x},t\right)\\
			&\dot{\Vec{x}} = \vec{f}_D\left(\vec{x},t\right)\\
			&\vec{x}\left(t_{0,j}\right) = \Vec{x}^{-}_{0,j}+\Vec{B}\vec{u}_{0,j}\\
			& || \vec{u}_{0,j} || \le u_{max}\\
			& t_{0,j} \le t_{h,j}\le T_{h,j}
		\end{aligned}
		\label{eq: Rechability analysis optimization}
	\end{equation}
	where $\vec{x}$ and $\vec{y}_i$ are respectively the spacecraft state in an inertial coordinate frame and in the abstract space associated with the region $\Omega_i$. The potentials $\omega_i$ and $V_C$ are continuous scoring functions indicating respectively the satisfaction of mission objectives and operational constraints. For the application proposed in this work only impact and escape constraints are considered. Given a pair $\{\vec{u}_{0j},t_{hj}\}$ the objective function and its gradient can be computing by integrating Eq.\eqref{eq: derivatives 4}.	Studying the behaviour of $ \dot{z}$ it is possible to perform a domain reduction by finding a criterion to define $t_{h,j} \left(\Vec{u}_{0,j},  \dot{z}\right)$. This allows to implement gradient based approaches following only the direction of $\boldsymbol{\eta}$. Three techniques are implemented: 1) Heuristic Sample Based (HSB), 2) Parallel Gradient Descent (PGD) and, 3) Neighbourhood Tree Exploration (NTE). The first one is the same proposed in \cite{rizza2023goal}, the second consists in performing an initial sampling on the control domain and then perform the gradient descent from a small subset of best scores, whilst NTE consists in alternating small size sample based exploration with gradient descent arcs for very few iterations.
		\begin{equation}
		\left\{ 
		\begin{aligned}
			& J = z(t_{h,j})  \\
			&\nabla J = \left[\dfrac{\partial J}{\partial t_{h,j}} , \dfrac{\partial J}{\partial \Vec{u}_{0,j}} \right]^T\ = \left[ \dot{z}\left( t_{h,j}\right),\boldsymbol{\eta} \left(t_{h,j} \right) \right]^T\\
			& \dot{z} = \dfrac{1}{T_{h,j}}\left(\sum_{i=1}^{n_{\Omega^A_i}} \omega_i \left(\Vec{y}_i \left(\Vec{x} \left(t\right),t\right)\right) +V_C \left(\vec{x} \left(t\right) \right) \right) \\
			& \dot{\Vec{x}} = \vec{f}_D\left(\vec{x},t\right)\\
			& \dot{\boldsymbol{\eta}} = \dfrac{1}{T_{h,j}}\left(\sum_{i=1}^{n_{\Omega_i}} \dfrac{\partial \omega_i}{\partial \Vec{y}_i}\dfrac{\partial \Vec{y}_i}{\partial \Vec{x}}  +   \dfrac{\partial V_C}{\partial \Vec{x}} \right) \STM{t}{t_{0,j}} \Vec{B} \\
			& \dot{\Phi}\left(t,t_{0,j} \right) = \dfrac{\partial \Vec{f}_D \left( \Vec{x},t\right)}{\partial \Vec{x}} \Phi\left(t,t_{0,j} \right)\\
			&\vec{x}\left(t_{0,j}\right) = \Vec{x}^{-}_{0,j}+\Vec{B} \vec{u}_{0,j}\\
			&z\left( t_{0,j}\right) = 0\\
			& \boldsymbol{\eta} \left( t_{0,j}\right) = \Vec{0}\\
			& \STM{t_{0,j}}{t_{0,j}} = \Vec{I}_{6\times6}
		\end{aligned}
		\right.
		\label{eq: derivatives 4}
	\end{equation}
	\section{Scenario definition}\label{sec: scenario definition}
\begin{figure}[h!]
		\centering
		\includegraphics[width=0.8\linewidth]{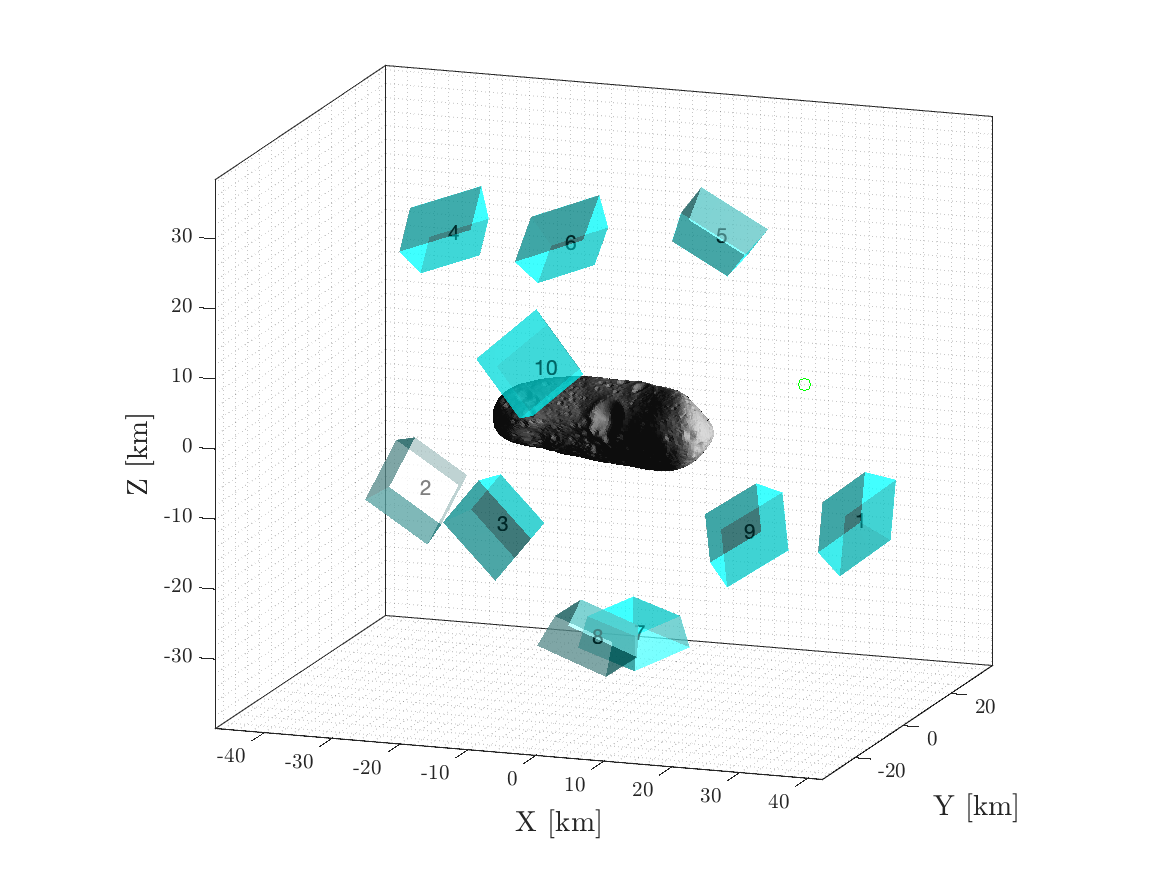}
		\caption{Asteroid 433 Eros with the selected set of observation regions.}
		\label{fig: observation regions}
	\end{figure}

	The proposed methodology is applied to the case of proximity operations about asteroid 433 Eros. A set of ten features is defined on the surface of the asteroid and mapped into observation regions as shown in Figure\ref{fig: observation regions}. Scientific requirements are defined in terms of range, off-nadir pointing and phase angle with respect to the features.
	\section{Results} \label{sec: results}
\begin{figure}[h!]
		\centering
		\includegraphics[width=1\linewidth]{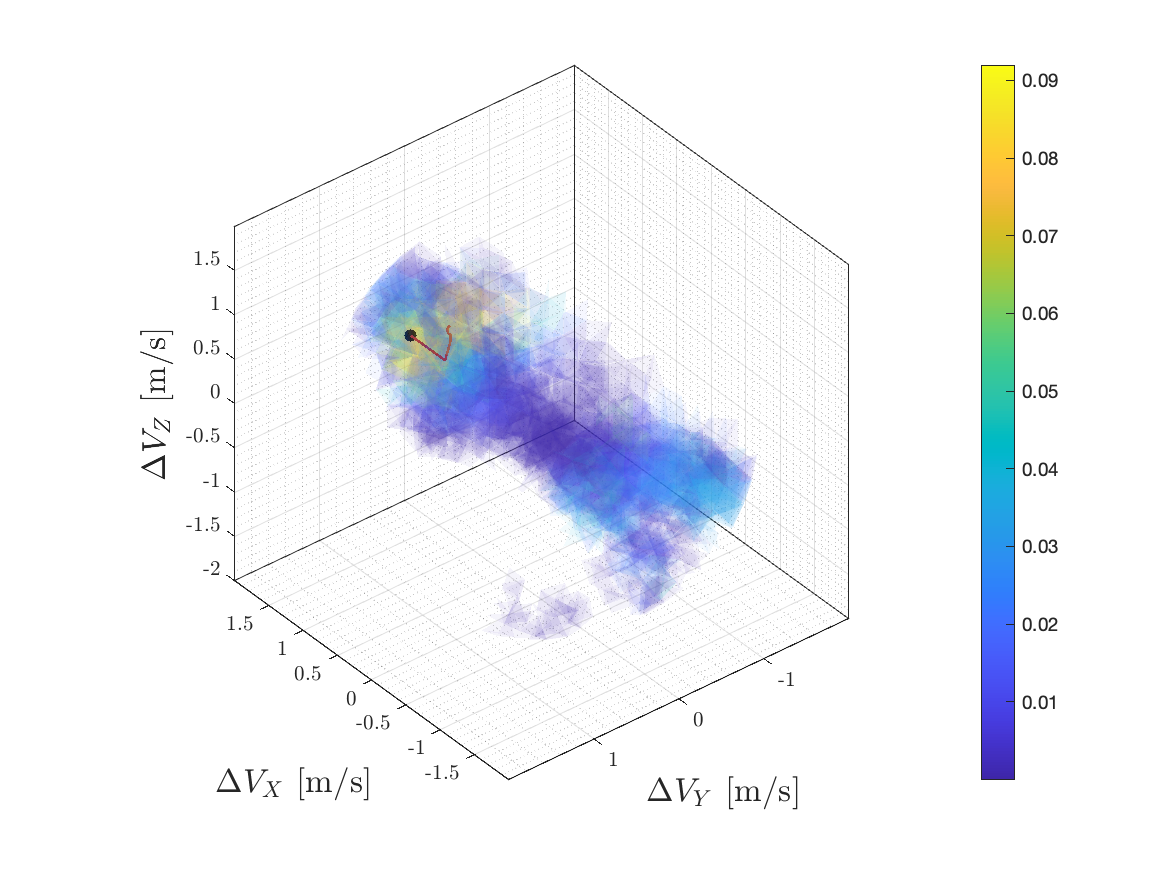}
		\caption{Example of reachable set exploration.}
		\label{fig: NTE_example}
	\end{figure}

\begin{figure}[h!]
		\centering
		\includegraphics[width=1\linewidth]{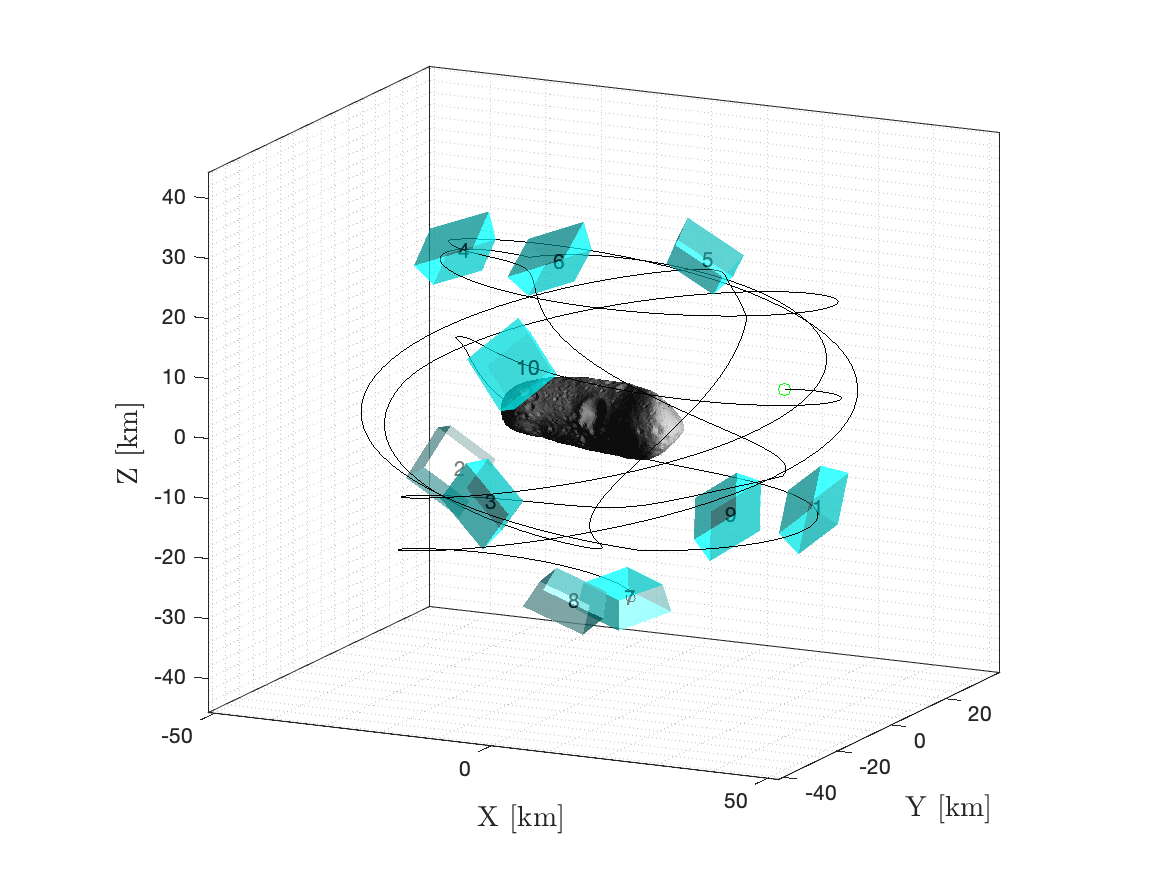}
		\caption{Optimized trajectory in asteroid fixed frame with gradient based NTE.}
		\label{fig: body_frame}
	\end{figure}

	Performances are compared with a Ground Truth (GT) generated with a Monte Carlo analyses. Tab\ref{tab: results} summarizes the achieved results in terms of function evaluations and final goal achieved.
	\begin{table}[h]
		\centering
		\caption{Performances and computational effort. }
		\begin{tabular}{ c c c } 
			\hline
			\hline
			& Fcn Eval & Goal Achievements \\ 
			\hline
			GT & 3516 & 100 \%\\ 
			NTE&157&80\%\\ 
			PGD& 287& 60\% \\ 
			HSB &140& 30\%\\						
			\hline
			\hline
		\end{tabular}
		\label{tab: results}
	\end{table}
	Results show that the gradient based NTE and PGD outperforms the simple sample based approach with a similar number of function evaluation. In particular, the switching between gradient descent and sampling ramification of the NTE reduces the probability of converging to a local minimum as often happens with the PGD. An example of the reachable set exploration with NTE is shown in Figure\ref{fig: NTE_example} with the colours indicating the value of the objective function. The full optimized trajectory in asteroid-fixed frame can instead  be observed in Figure\ref{fig: body_frame}.
	
	\bibliography{sample}
	
	%{\small
		%
		%\begin{thebibliography}{99}
		%
		%
		%\bibitem{scheeres2016orbital}
		%Scheeres, Daniel J, {\it Orbital motion in strongly perturbed environments: applications to asteroid, comet and planetary satellite orbiters}, %% Capitalize book titles
		%2016, pp. 18-33, Springer, Chichester, UK.
		%
		%\bibitem{de2010taking}
		%De Novaes Kucinskis, Fabr{\'\i}cio and Ferreira, Maur{\'\i}cio Gon{\c{c}}alves Vieira, {\it Taking the ECSS autonomy concepts one step further}, %% Capitalize book titles
		%SpaceOps 2010 Conference “Delivering on the Dream” Hosted by NASA Mars, pp. 25-30, 2010
		%
		%\bibitem{surovik2016autonomous}
		%De Novaes Kucinskis, Fabr{\'\i}cio and Ferreira, Maur{\'\i}cio Gon{\c{c}}alves Vieira, {\it Taking the ECSS autonomy concepts one step further}, %% Capitalize book titles
		%SpaceOps 2010 Conference “Delivering on the Dream” Hosted by NASA Mars, pp. 25-30, 2010
		%
		%
		%
		%%% You may use other variants if necessary.
		%
		%\end{thebibliography}}

	\end{document}